\documentclass[journal]{IEEEtran}

\usepackage{graphicx} 
\usepackage{amsmath} 
\usepackage{color,soul} 
\usepackage{xcolor}
\usepackage{amssymb} 
\usepackage{textcomp} 
\DeclareUnicodeCharacter{2212}{\textminus}
\usepackage{cite} 
\usepackage{booktabs} 
\usepackage{subcaption} 
\usepackage{xurl}
\usepackage[hidelinks]{hyperref}
\usepackage{enumitem}
\usepackage{bm}
\usepackage{mathtools}
\usepackage{accents}
\usepackage{caption}
\usepackage{bm}
\usepackage{booktabs}
\usepackage{units}
\usepackage{diagbox}
\usepackage{multicol}
\usepackage{multirow}
\usepackage{balance}

\usepackage{amsmath,amsfonts,amsthm,amssymb}
\usepackage[mathscr]{euscript}
\usepackage[bb=boondox]{mathalfa}
\usepackage[algoruled]{algorithm2e}
\usepackage{array}
\usepackage{mdframed}
\usepackage{bm}

\usepackage[noabbrev,capitalize]{cleveref}

\crefname{equation}{}{}
\Crefname{equation}{}{}
\crefname{section}{}{}
\Crefname{section}{}{}

\theoremstyle{definition} 

\theoremstyle{plain} 

\theoremstyle{remark} 

\usepackage{tikz}

\DeclareMathOperator*{\argmin}{arg\,min}

\makeatletter
\newcommand{\pushright}[1]{\ifmeasuring@#1\else\omit\hfill$\displaystyle#1$\fi\ignorespaces}
\newcommand{\pushleft}[1]{\ifmeasuring@#1\else\omit$\displaystyle#1$\hfill\fi\ignorespaces}
\makeatother

\newcolumntype{C}[1]{>{\centering\arraybackslash}p{#1}} 

\newcounter{box}

\title{Strategic and Grid-Aware Maintenance Planning\newline of Offshore Wind Farms}
\author{Robert Mieth\textsuperscript{1,*}, Ahmed Aziz Ezzat\textsuperscript{1}

\thanks{\textsuperscript{1}Department of Industrial and Systems Engineering, Rutgers University.}
\thanks{\textsuperscript{*}Corresponding author.  robert.mieth@rutgers.edu.}
}

\begin{document}

\maketitle

\begin{abstract}
    Wind turbines require regular maintenance and the resulting costs are a substantial component of a wind farm's cost of electricity production. As a result, there has been ongoing interest in improving wind turbine maintenance scheduling to find an optimal balance between maintenance costs and the risk of failure or unplanned repairs. What remains understudied is the opportunity for wind farms to schedule maintenance in the context of grid conditions and electricity market clearing.
    This paper contributes to closing this gap by modeling and studying wind farm maintenance planning in a grid and electricity market context. We also review current U.S. practice of wind farm maintenance scheduling, which motivates this paper and its models.
    Our focus is the derivation of a strategic maintenance planning problem, alongside an efficient solution approach, in which the wind farm operator aims to submit a derated wind farm capacity such that the resulting market clearing and electricity prices maximize its profit while ensuring that all required maintenance can be performed. We derive and study both deterministic and stochastic versions of the model, with the latter considering environmental and operational uncertainties.
    We conduct numerical experiments using the IEEE RTS 96-bus testbed with real-world offshore wind farm data and investigate the roles of farm and turbine size, as well as forecast quality. We observe an alignment between strategic and grid-serving maintenance planning and find that wind farms can improve their bottom line with low impact on system costs.
\end{abstract}



\section{Introduction}

Offshore wind farms are an economically competitive and clean energy source. 
In Europe, already 37~GW of offshore wind capacity are connected to the grid \cite{windeurope2025stats}.
In the U.S., there is currently around \unit[1]{GW} of offshore wind capacity in operation with another roughly \unit[5]{GW} in the construction or commissioning pipeline. 
Despite current headwinds in the U.S. offshore wind industry, the 2022 \unit[4.37]{billion} dollar valuation of the New York Bight offshore wind lease areas \cite{DOI2022NewYorkBight} remains a strong signal for future offshore wind capacity, especially in the context of data center-driven energy demand growth.

In the absence of fuel cost, operation and maintenance (O\&M) cost are the main contributor to offshore wind running and variable costs \cite{stehly20212020}. 
Compared to onshore wind turbines, these costs are amplified offshore by the relatively larger scale of the turbines, the requirement of specialized maintenance ships, and access uncertainty due to stochastic atmospheric and oceanic (``metocean'') conditions \cite{petersen2026accessibility}. 
At the same time, offshore wind farms can have a greater impact on grid operations in case of planned or unplanned outages due to their larger generating capacity and spatial density, potentially posing grid reliability risks when multiple contiguous wind farms experience large-scale generation losses.
Hence, there is value for the grid operator to coordinate wind farm maintenance, similar to existing generator outage scheduling.
Naturally, there is also an opportunity for wind farm operators to strategically manage their maintenance needs based on anticipated electricity market conditions.
To the authors' knowledge, no analyses and modeling efforts have addressed this latter opportunity. 
This paper aims to close this gap.

\subsection{Generator maintenance planning}

\begin{table*}
\centering
\caption{Maintenance Outage Reporting Requirements Across ISOs/RTOs}
\label{tab:derate_requirements}
\begin{tabular}{ l  p{3.2cm}  l  p{2.8cm}  l | c}
\toprule
\textbf{ISO/RTO} & \textbf{Lead time} & \textbf{Aggregation} & \textbf{Threshold} & \textbf{Reference} & \textbf{Installed Wind} \\
\midrule
\textbf{CAISO} 
    & 8 days (``planned maint.'')\newline 4 days (``unplanned maint.'')  & Farm-level & \unit[1]{MW}
& \cite[Sec.~9.3.6.4.1(a)]{CAISO2024Tariff} & \unit[8.3]{GW} \cite{CAISO_Key_Statistics_2025}\\
\textbf{ERCOT} 
    & 3 days\textsuperscript{*} (planned/maint.) & Farm-level & Not specified & \cite[Section 3.1.6]{ercot2025nodal} & \unit[40]{GW} \cite{ERCOT_Monthly_Report_July_2025}\\ 
\textbf{ISO-NE} 
    & 15 days (planned)\newline None\textsuperscript{\textdagger} (maintenance)  & Farm-level & \unit[10]{MW} or \unit[2]{\%}  capacity &  \cite{ISO-NE_MR1_AppendixA_2025,ISO-NE_OP5_2025} & \unit[1.7]{GW}  \cite{ISO_NE_Wind_Integration} \\
\textbf{MISO}
    & 24 months (planned)\newline None (maintenance) & Farm-level & Any & \cite[Section 4.1]{miso2024outage} & \unit[30]{GW} \cite{MISO_WindSolarCredit_2024} \\ 
\textbf{NYISO}
    & 2 days\textsuperscript{\textdaggerdbl} (planned/maint.) & Farm-level & \unit[1]{MW} lasting $\ge$\unit[1]{h} & \cite[Section 4]{nyiso2024windsolar} & \unit[2]{GW} \cite{NYISO_GoldBook_2023} \\
\textbf{PJM}
    & 30 days (planned)\newline3 days (maintenance) & Farm-level & \unit[1]{MW} lasting $\ge$\unit[1]{h}\newline
    or \unit[10]{\%} capacity & \cite[Section 2.3]{PJMM10} & \unit[12.2]{GW} \cite{monitoringanalytics2024som_pjm_sec12}  \\
\textbf{SPP}
    & 14 days (planned)\newline None (maintenance) & Farm-level & \unit[10]{MW} lasting $\ge$\unit[30]{min} & \cite[Section 3]{spp_rc_outage_coordination_v32} & \unit[34.8]{GW} \cite{spp_annual_state_of_the_market_2024} \\
\bottomrule
\end{tabular}\\
\flushleft
\vspace{-0.5em}
\hspace{4em} 
\begin{minipage}{0.86\textwidth}
\textsuperscript{*}Due to wind farms being ``non-reliability'' resources; The 3 day lead time is valid for Level III maintenance (required within 90 days to prevent failure). Level II (required within 30 days) and Level I (required within 24 hours) require 2 days and no lead time, respectively.\\
\textsuperscript{\textdagger}Ideally 1 day; Maintenance must begin within 14 days.\\
\textsuperscript{\textdaggerdbl}Wind-specific requirements 
\end{minipage}
\end{table*}

Almost every asset in the power system requires regular maintenance to prevent unplanned (forced) outages and equipment damage, which are typically more costly than planned maintenance outages. 
Scheduling planned generator outages, alongside the maintenance of other critical equipment like transmission and distribution assets, in power systems with high reliability requirements is challenging and much research has focused on this topic.
This includes classic reliability considerations in vertically integrated power systems, predating the power industry's deregulation and the resulting complication of coordinating maintenance outages between the grid operator and third-party generation owners \cite{billinton2005composite}. 
With increasing computational ability, also more complex models to compute optimal (vertically integrated) maintenance windows based on a cost/benefit analysis were proposed \cite{abiri2012optimized}.

In the context of competitive electricity markets, generator maintenance was studied through the lens of finding a consensus between one or multiple generation companies and the grid operator.
Conjeo \textit{et al.} \cite{conejo2005generation} make an early proposal for an incentive scheme that guides generators to align their maintenance schedule with a reliability-maximizing grid objective.
Pand\v{z}i\'{c}~\textit{et al.} \cite{pandzic2012epec} model and analyze the outage scheduling problem as an equilibrium problem with profit-maximizing, strategic behavior of the generators.   
Wang~\textit{et~al.}~\cite{wang2016coordination}, on the other hand, discuss a welfare-maximizing bid-based coordination scheme.
We also refer to the review in \cite{froger2016maintenance} for a detailed collection of research in both vertically integrated and deregulated power systems.

The maintenance scheduling problems in \cite{billinton2005composite,conejo2005generation,pandzic2012epec,wang2016coordination,abiri2012optimized,froger2016maintenance} focus on the cost of generator operations, the cost of grid operations, or the reliability of grid operations based on fixed and pre-defined maintenance requirements. 
A more recent line of research combines operation-aware and sensor data-driven generator degradation models with optimal maintenance scheduling.
This line is spearheaded by the works of Yildirim \textit{et al.} \cite{yildirim2016sensor,yildirim2019leveraging} and Basciftci \textit{et al.} \cite{basciftci2018stochastic,basciftci2020data}, where Yildirim \textit{et al.} \cite{yildirim2016sensor} propose an operations-informed sensor-driven approach to generator maintenance planning. 
Basciftci \textit{et al.} \cite{basciftci2018stochastic} propose a maintenance planning approach that internalizes probabilistic equipment degradation models, and \cite{basciftci2020data,yildirim2019leveraging} provide an extension that also considers the contribution of generator utilization to equipment degradation.
Rokhforoz \textit{et al.} \cite{rokhforoz2021multi} extend the idea of sensor data-driven predictive maintenance planning with the necessity of grid coordination and propose a bid-based mechanism in the spirit of \cite{wang2016coordination,pandzic2012epec}.

It is noteworthy that renewable energy remains largely a side note in the generator maintenance planning literature.
Some works consider the role of renewables in grid operations. For example, Randall \textit{et al.} \cite{randall2025risk} extend the operation-aware maintenance scheduling problem from \cite{basciftci2020data} to internalize uncertainty from renewable injections. 
Related, but without predictive capabilities, Sadeghian \textit{et al.} \cite{sadeghian2021risk}, consider the risk from an uncertain, renewable-driven need for flexible resources in the maintenance scheduling of thermal power plants. 
An emerging body of literature, which we review in the following section, studies the maintenance needs of wind power generators, but so far, analyses of explicitly coordinating such maintenance with the grid remain understudied.

\subsection{(Offshore) Wind farm maintenance}

We focus our discussion on offshore wind farms due to their larger scale and greater operational impact, and because the offshore wind maintenance-planning problem encompasses all layers of complexity that are also relevant for onshore wind.
These complexity layers comprise (i) uncertain maintenance opportunity cost due to weather-dependent production, (ii)~uncertain site-access due to unsafe wind and/or metocean conditions, and (iii) complexity of maintenance operations due to specialized transport vessel and crew requirements \cite{papadopoulos2024stochos}.
The central motif of the offshore wind maintenance literature to date is the optimal balance of the cost of preventive maintenance and the risk of delaying maintenance with the goal of reducing O\&M cost per produced megawatt-hour of wind energy. 
See Besnard \textit{et al.} \cite{besnard2009optimization}, Shafiee~\textit{et al.} \cite{shafiee2015opportunistic}, Yildirim \textit{et al.} \cite{yildirim2017integrated}, and Papadopolous \textit{et al.} \cite{papadopoulos2021seizing,papadopoulos2023joint,papadopoulos2024stochos}.

The explicit relationship between offshore wind maintenance planning and grid operations is understudied. 
The above-mentioned methods \cite{yildirim2017integrated,papadopoulos2024stochos} consider electricity prices to optimize the opportunity cost of maintenance, but do not otherwise consider grid operations. 
Donnelly \textit{et al.} \cite{donnelly2025silver} offer the first maintenance planning strategy that considers curtailment signals from the grid operator,  followed by \cite{wang2026gridinformed} who formalized this grid-farm interaction concept into a rigorous maintenance optimization model that extends Donnelly \textit{et al.}'s rule-based approach. 
Otherwise, literature on wind farm outages and their impact on the grid remains sparse. 
Our previous work in Mieth \textit{et al.} \cite{mieth2022risk} considers unplanned wind farm outages on contingency reserve schedules building on a wind reliability model from~\cite{sulaeman2016wind}.

\subsection{Opportunities from current grid operations}
\label{ssec:opportunities_from_grid_operations}

Effective and grid-aware offshore wind maintenance planning also depends on the requirements set by system operators\footnote{We stick to using the term ``system operators'' in this paper to avoid having to make the technically correct but more clunky distinction between ISOs and RTOs.}, which we review in this section and derive two central opportunities that inform the central motivation and modeling of this paper. 
To this end, we studied the system operators' public-facing market and generator operation rules to identify the requirements for wind farm operators to report maintenance-related outages. 
Table~\ref{tab:derate_requirements} summarizes our findings focusing on the required lead time to report capacity derates, the granularity of reporting, and the threshold at which a capacity derate becomes mandatory to report in the U.S. 
Table~\ref{tab:derate_requirements} also shows currently installed wind capacity for reference and we highlight that there is offshore wind potential in the service territories of all system operators.

While the requirements are heterogeneous across system operators, a few similarities emerge. 
Most system operators recognize ``variable'' or ``intermittent'' energy resources as a distinct generation technology group, but generally do not enforce outage reporting requirements that differ from other conventional generators. 
Compared to the reporting threshold for derates of conventional generators, CAISO, NYISO, and PJM, enforce a lower derate reporting threshold of \unit[1]{MW} for wind farms with the goal of maintaining accurate wind power forecasts. 
Also, all system operators only require farm-level reporting of derates. This means wind farm operators are not required to report outages of individual turbines, but only the total effective wind farm capacity as the sum of the nameplate capacities of all available turbines.

The required reporting lead time varies across system operators and depends on the type of outage and, as shown in Table~\ref{tab:derate_requirements}, on whether an outage is considered a ``planned'' outage or a ``maintenance'' outages. 
Although exact definitions of these outage types differ across system operators, they generally follow the same structure: Planned outages or derates are motivated by regular maintenance intervals and not by specific reliability concerns. 
Maintenance outages or derates, on the other hand, are preventive actions triggered by credible elevated generator outage risks, for example indicated by processed sensor data. 
Most system operators exclude wind farms from reliability considerations and therefore only enforce reporting lead times for planned outages on the order of days or weeks. 
The lead time for reporting derates related to reliability-triggered maintenance is generally on the order of a few days across all system operators.
The preventive maintenance methods \cite{shafiee2015opportunistic,yildirim2017integrated,papadopoulos2021seizing,papadopoulos2023joint,papadopoulos2024stochos} discussed above all base their maintenance schedules on an expected-time-to-failure (or similar) metric using sensor data and/or detailed turbine models. 
We therefore argue, that the preventive maintenance proposed in \cite{shafiee2015opportunistic,yildirim2017integrated,papadopoulos2021seizing,papadopoulos2023joint,papadopoulos2024stochos} would fall under the reporting requirements for ``maintenance''. 

Regardless of the exact categorization, the following two opportunities emerge from the requirements summarized in Table~\ref{tab:derate_requirements}: (i) Optimal preventive maintenance actions can be scheduled at a relatively short notice, allowing wind farm operators to take into account short-term forecasts on weather, metocean conditions, electricity prices, and other grid conditions. 
(ii) Turbine outages only need to be reported as farm-level derates, allowing wind farm operators to freely assign turbine outages in derate windows. 
The modeling and discussion in this paper takes advantage of these opportunities. 

\subsection{Objective and Contributions}


Motivated by the above discussion, the goal of this paper is to model and analyze opportunities for operational wind farm maintenance scheduling in the context of grid conditions and electricity market clearing.
To this end, we formulate a wind farm maintenance planning model that is informed by recent literature on detailed wind farm operations (Section~\ref{sec:basic_model_formulation}). 
We then embed this model into an economic dispatch market clearing problem and formulate a bilevel, strategic version where the wind farm operator anticipates the grid operator's response to the wind farm capacity derate that is necessary to perform required maintenance. 
To the best of our knowledge, this is the first study to formulate such a strategic maintenance planning model for wind farms. 
We then extend the formulation to consider uncertainty in wind power production and access to the site (Section~\ref{sec:maint_planning_under_uncertainty}).
We use these models to perform comprehensive numerical experiments and analyze the roles of farm capacity, turbine size, as well as forecast quality (Section~\ref{sec:case_study}).

\section{Deterministic Model Formulation}
\label{sec:basic_model_formulation}

We first present deterministic versions of the wind operator's and grid operator's problems. We use these models to set up a strategic maintenance planning problem alongside a tractable solution approach.
For the modeling and analysis goals of this paper we model one representative offshore wind operator that we assume has very good (perfect) knowledge of the grid operator and its decision-making. 
We further discuss this assumption below and also refer to the discussion in \cite{wang2017look}.

\subsection{Wind farm operator: Maintenance scheduling}

We consider an offshore wind farm operator with $I$ wind turbines that require regular maintenance.
For each wind turbine $i\in[I]$, the below optimization problem \eqref{eq:basic_maintenance_model} finds a profit-maximizing maintenance schedule $\{m_{i,t}\}_{t=1}^T$ over a horizon of $T$ timesteps (e.g., hours) that is divided in $D$ periods (e.g., days). For example, for one week at a one-hour resolution, we have $T=7\times 24=168$ hours and $D=7$ days.
\begin{subequations}
\begin{align}
\max_{p_t^w, m_{i,t}} \quad 
    & \sum_{t\in[T]}\Big(p^{\rm w}_{t}\lambda_{t} 
    - \sum_{i\in[I]}m_{i,t} K_t^{\rm var}\Big) -\sum_{d\in[D]}mp_{d} K^{\rm fix} \hspace{-1cm}\\
\text{s.t.} \quad 
    & 0\le p^{\rm w}_{t} \le \sum_{i\in[I]}\overline{p}_{i}(1-m_{i,t})A_{i,t} && \forall t \label{eq:power_from_derated_capacity}\\ 
    & \sum_{t\in[T]}m_{i,t} \ge M^{\rm req}_i &&\forall i \label{eq:maintenance_requirement}\\
    & ms_{i,t} \ge m_{i,t} - m_{i,t-1} && \forall t \label{eq:maintenance_start_indicator}\\
    & \sum_{s=t-(M^{\rm int}_i-1)} ^t ms_{i,s} \le m_{i,t} && \forall t \label{eq:minimum_maintenance_time} \\
    & \sum_{i\in[I]}m_{i,t} \le M^{\rm par}_t && \forall t \label{eq:max_parallel_maintenance}\\
    & mp_{d} \ge \sum_{i\in[I]} m_{i,t} && \hspace{-2.5cm}\forall t \in [T]_d,\ \forall d\in[D] \label{eq:parallel_maintenance_indicator} \\
    & m_{i,t} \in \{0,1\}, \ ms_{i,t} \ge 0 && \forall i,t. \label{eq:maintenance_binary}
\end{align}%
\label{eq:basic_maintenance_model}%
\end{subequations}%
Note that the variables underneath the maximization operator reflect the collection of primary decision variables over their respective index sets. 
We omit listing auxiliary variables for readability.

The wind farm operator aims to maximize their profit composed of revenue from selling power to the grid and cost of maintenance.
At each time step $t$ the wind farm injects a total power $p_{t}^{\rm w}$ into the grid at price $\lambda_t$ to produce profit $p^{\rm w}_{t}\lambda_{t}$. We write $\{1,...,T\}=[T]$ and use $[T]_d$ as a shorthand for the index set of the timesteps that are in period $d$.
Profit is reduced by ``fixed'' ($K^{\rm fix}$) and ``variable'' ($K^{\rm var}$) maintenance cost. Fixed costs capture per-period costs that are independent of the actual number of maintenance intervals executed in this period, for example cost of renting a service vessel.
Variable costs are incurred per maintenance slot, for example hourly personnel wages, equipment rental, and materials.
Binary variables $m_{i,t}\in\{0,1\}$ indicate whether turbine $i$ is out for maintenance at time $t$ and variable $mp_d$ collects the highest number of parallel maintenance during period $d$ as per \cref{eq:parallel_maintenance_indicator}.
Constraint~\cref{eq:power_from_derated_capacity} limits wind power injection to the turbine capacity $\overline{p}_{i}$ times an availability factor $A_{i,t}\in[0,1]$ if $m_{i,t}=0$, i.e., the turbine is \textit{not} out for maintenance.
The per-period turbine-specific availability factor$A_{i,t}\in[0,1]$ (or ``capacity factor''), depends on wind conditions. 
Constraint \cref{eq:maintenance_requirement} enforces that the required number of maintenance periods $M_i^{\rm req}$ is achieved within the horizon $T$ for a turbine.
This requirement is informed by the number of maintenance tasks the farm operator needs to perform during the planning horizon, for example as informed by a statistical remaining-useful-life assessment or a condition-based monitoring system.
For the sake of the analyses in this paper, we assume that no unexpected failures occur during the planning horizon, provided that all scheduled maintenance requirements have been successfully completed.
Constraint \cref{eq:maintenance_start_indicator} defines the starting time of a maintenance interval, i.e., $ms_{i,t}=1$ at the first timestep of a sequence of consecutive timesteps with maintenance, and constraint \cref{eq:minimum_maintenance_time} enforces a minimum number $M^{\rm int}_i$ of consecutive time steps for a maintenance interval.
Constraint \cref{eq:max_parallel_maintenance} restricts the number of parallel maintenance at time $t$ to $M^{\rm par}_t$.
Parameters $M^{\rm int}_i$, $M^{\rm par}_t$, $K^{\rm var}$, and $K^{\rm fix}$ are informed by practical considerations for maintenance crew scheduling and availability (e.g., \cite{dinwoodie2015reference}).

\subsection{System operator: Economic dispatch}

At each time step, the power system operator computes a set of cost-minimal power injections $p_{g,t}^{\rm c}$ from $G$ controllable generators and available wind production to meet a demand $D_t$.
The system operator uses a (potentially derated) wind farm capacity $p_t^{\rm F}$ and a farm-wide availability factor $A_t$.
For a given maintenance schedule, the effective farm capacity is 
\begin{equation*}
    p_t^{\rm F} = \sum_{i\in[I]}\overline{p}_{i}(1-m_{i,t}).
\end{equation*}
The resulting dispatch problem of the system operator is:
\begin{subequations}
\begin{align}
\min_{p_{g,t}^{\rm c}, p_t^{\rm w}} \quad 
    & &&\sum_{t\in[T]}\sum_{g\in[G]} c_g p^{\rm c}_{g,t} \\
\text{s.t.} \quad 
    &(\lambda_t):\ && \sum_{g\in[G]} p_{i,t}^{\rm c} + p_t^{\rm w} = D_t && \forall t \label{eq:ed_enerbal}\\
    &(\underline{\mu}^{\rm c}_{g,t},\overline{\mu}^{\rm c}_{g,t}):\   &&0 \le p_{g,t}^{\rm c} \le p^{\rm c,max}_{g} && \forall g,t \label{eq:ed_generation_cap} \\
    &(\underline{\mu}^{\rm w}_t,\overline{\mu}^{\rm w}_t):\  &&0 \le p_t^{\rm w} \le p_t^{F} A_t && \forall t. \label{eq:ed_wind_cap}
\end{align}%
\label{eq:ed_problem}%
\end{subequations}%
The system operator aims for a cost-minimal economic dispatch by minimizing the cost of generation determined by the generators' cost parameters $c_g$.
We assume that wind production enters the economic dispatch at zero marginal cost, consistent with the near-zero short-run marginal cost of wind generation and common market bidding practices \cite{pjm2022intermittent}.
Balance between production and demand is enforced by the power balance in \cref{eq:ed_enerbal}.
Production of controllable generators is limited by their installed capacity $p^{\rm c,max}_{g}$ in constraint \cref{eq:ed_generation_cap}.
Constraint \cref{eq:ed_wind_cap} limits the possible power injection from wind to the total farm capacity multiplied by the availability factor $A_t$.
Variables in parentheses denote dual variables of the respective constraints, which we will utilize for derivations below.
The dual of the energy balance ($\lambda_t$) sets the price for a unit of power at time~$t$.

\subsection{Grid-serving maintenance planning}

The maintenance planning model in \eqref{eq:basic_maintenance_model} assumes a fixed set of prices and the economic dispatch in \cref{eq:ed_problem} assumes a fixed wind farm capacity. 
A straightforward alternative would be a wind farm maintenance schedule that minimizes cost from the perspective of the grid operator.
We call this \textit{grid-serving} maintenance planning modeled as:
\begin{subequations}
\begin{align}
\min \quad 
    &\sum_{t\in[T]}\sum_{g\in[G]} c_g p^{\rm c}_{g,t} \\
\text{s.t.} \quad 
    &\text{Grid constraints: \cref{eq:ed_enerbal,eq:ed_generation_cap}} \\
    & \text{Maintenance constraints: \cref{eq:maintenance_binary,eq:maintenance_requirement,eq:max_parallel_maintenance,eq:minimum_maintenance_time,eq:maintenance_start_indicator,eq:power_from_derated_capacity,eq:parallel_maintenance_indicator}},
\end{align}
\label{eq:grid_serving_maintenance}
\end{subequations}
where $A_{i,t}=A_t,\ \forall i$.

\subsection{Deterministic strategic maintenance scheduling}

A second alternative is that the wind farm operator anticipates the impact of its wind farm capacity derates on prices and and dispatch decisions. 
To maximize profit, the wind farm operator naturally prioritizes maintenance at times when price $\lambda_t$ and availability $A_{i,t}$ is low. 
High levels of wind availability may drive prices down, and vice versa, creating an opportunity for the wind farm operator to strategically derate wind farm capacity such that all maintenance constraints can be met while profit is maximized. 

\subsubsection{Model formulation}
The resulting strategic maintenance scheduling problem is the following bi-level problem:
\allowdisplaybreaks
\begin{subequations}
\begin{align}
\max_{p_t^{\rm F}} \quad 
    & \sum_{t\in[T]}\Big(p^{\rm w}_{t}\lambda_{t} 
    - \sum_{d\in[D]}mp_{d} K^{\rm fix}
    - \sum_{i\in[I]}m_{i,t} K_t^{\rm var}\Big) \label{eq:bilevel_objective} \\
\text{s.t.} \quad 
    & 0 \le p^{\rm F}_{t} \le \sum_{i\in[I]} \overline{p}_{i}(1-m_{i,t}), \quad \forall i,t \label{eq:derated_capacity_bilevel}\\
    &\text{\cref{eq:maintenance_binary,eq:maintenance_requirement,eq:max_parallel_maintenance,eq:minimum_maintenance_time,eq:maintenance_start_indicator,eq:parallel_maintenance_indicator}} \label{eq:bilevel_upper_constraints}\\
    & \lambda_t, p^{\rm w}_t \in \Big\{ \argmin \sum_{t\in[T]}\sum_{g\in[G]} c_g p^{\rm c}_{g,t} \label{eq:bilevel_lower_objective}\\ 
    & \hspace{5em} \text{ s.t.   \cref{eq:ed_enerbal,eq:ed_generation_cap,eq:ed_wind_cap}} \Big\} \label{eq:bilevel_lower_constraints}
\end{align}%
\label{eq:bilevel_main}
\end{subequations}%
\allowdisplaybreaks[0]%
The upper level wind operator (leader) problem given by \cref{eq:bilevel_objective,eq:derated_capacity_bilevel,eq:bilevel_upper_constraints}, decides on the optimal, potentially derated, wind warm capacity $p_t^{F}$ to submit to the system operator, who then computes prices $\lambda_t$ and dispatch schedules $p_t^{\rm w}$ as in \cref{eq:ed_problem}. This lower level system operator (follower) problem is given by \cref{eq:bilevel_lower_constraints,eq:bilevel_lower_objective}.
Note that as for \eqref{eq:grid_serving_maintenance}, we assume here consistency in the wind power availability across turbines and between the wind operator and the system operator, i.e., $A_{i,t}=A_t,\ \forall i$.



\subsubsection{Solution approach}
\label{ssec:solution_approach}

To solve  \cref{eq:bilevel_main} we reformulate it as a tractable single-level equivalent problem by replacing the lower level problem with its first-order optimality (KKT) conditions.

The Lagrangian of \cref{eq:ed_problem} is:
\begin{equation}
\begin{aligned}
L = &\sum_{t\in[T]} \sum_{g\in[G]} c_g p_{g,t}^{\rm c} - \sum_{t\in[T]}\lambda_t(\sum_{g\in[G]} p_{g,t}^{\rm c} + p_t^{\rm w} - D_t)\\
& +  \sum_{t\in[T]} \overline{\mu}_t^{\rm w}(p_t^{\rm w} - p_t^{\rm F} A_t) - \sum_{t\in[T]} \underline{\mu}_t^{\rm w} p_t^{\rm w} \\
& + \sum_{t\in[T]} \sum_{g\in[G]} \overline{\mu}_{g,t}^c(p_{g,t}^c - p^{\rm c,max}_g) - \sum_{t\in[T]} \sum_{g\in[G]} \underline{\mu}_{g,t}{p}_{g,t}^c,
\end{aligned}
\end{equation}
leading to the dual objective of \cref{eq:ed_problem} given as
\begin{equation}
\max \sum_{t\in[T]} \big(\lambda_t D_t -  \overline{\mu}_t^w p_t^F A_t - \sum_{g\in[G]} \overline{\mu}_{g,t}^c \overline{p}_g\big),
\label{eq:dual_objective}
\end{equation}
and its KKT conditions
\allowdisplaybreaks
\begin{subequations}
\begin{align}
    & c_g -\lambda_t + \overline{\mu}_{t,g}^{\rm c} - \underline{\mu}_{t,g}^{\rm c} = 0 && \forall g,t \label{eq:kkt_bal_c}\\
    & -\lambda_t + \overline{\mu}_{t}^{\rm w} - \underline{\mu}_{t}^{\rm w} = 0 && \forall t \label{eq:kkt_bal_w}\\
    & \sum_{g\in[G]} p_{i,t}^{\rm c} + p_t^{\rm w} - D_t = 0 && \forall t \label{eq:kkt_enerbal}\\
    & p_t^{\rm w} - p_t^{\rm F} A_t \le 0 && \forall t \label{eq:kkt_wind_primal_upper}\\
    & p_{g,t}^{\rm c} - p^{\rm c,max}_{g} \le 0 && \forall g,t \label{eq:kkt_gen_primal_upper}\\
    & \overline{\mu}_{t}^{\rm w}(p_t^{\rm F} A_t - p_t^{\rm w}) = 0 && \forall t \label{eq:wind_comp_upper}\\
    & \overline{\mu}_{g,t}^{\rm c}(p^{\rm c,max}_{g} - p_{g,t}^{\rm c}) = 0 && \forall g,t \label{eq:gen_comp_upper}\\
    & \underline{\mu}_{t}^{\rm w} p_{t}^{\rm w} = 0 && \forall t \label{eq:wind_comp_lower}\\
    & \underline{\mu}_{g,t}^{\rm c} p_{g,t}^{\rm c} = 0 && \forall g,t\label{eq:gen_comp_lower}\\
    & p_{g,t}^{\rm c}, p_{t}^{\rm w}, \overline{\mu}_{g,t}^{\rm c}, \underline{\mu}_{g,t}^{\rm c}, \overline{\mu}_{t}^{\rm w}, \underline{\mu}_{t}^{\rm w}  \ge 0 \label{kkt:var_nonneg}\\
    &\lambda_t: \text{ free} \label{kkt:var_free}.
\end{align}
\end{subequations}
\allowdisplaybreaks[0]

\subsubsection{Resolving bilinear terms}

Consider the bilinear first term in \cref{eq:bilevel_objective}. From $\cref{eq:kkt_bal_w}$ we get
\begin{equation*}
    \lambda_t = \overline{\mu}_t^w - \underline{\mu}_t^w 
\end{equation*}
leading to 
\begin{equation*}
    \sum_t p_t^w \lambda_t = \sum_t (p_t^w\overline{\mu}_t^w - p_t^w\underline{\mu}_t^w).
\end{equation*}
From \cref{eq:wind_comp_lower} and \cref{eq:wind_comp_upper} we get, respectively:
\begin{align*}
    &p_t^w\underline{\mu}_t^w = 0 \\
    &p_t^w\overline{\mu}_t^w = \overline{\mu}_t^w\tilde{p}_t^FA_t.
\end{align*}
Because the lower level problem is linear, strong duality holds and we can equalize the primal and dual objectives of \cref{eq:ed_problem}:
\begin{equation*}
    \sum_t \sum_g c_g p_{g,t}^c = 
    \sum_t \big(\lambda_t D_t -  \overline{\mu}_t^w \tilde{p}_t^F A_t - \sum_g \overline{\mu}_{g,t}^c \overline{p}_g\big),
\end{equation*}
which we can use this to rewrite the first term of \cref{eq:bilevel_objective} as:
\begin{align*}
\sum_t p_t^w \lambda_t 
  &= \sum_t \overline{\mu}_t^w \tilde{p}_t^F A_t  \\
  &= -\sum_t \Big(\sum_g c_g p_{g,t}^c - \lambda_t D_t + \sum_g \overline{\mu}_{g,t}^c \overline{p}_g\Big).
\end{align*}

We rewrite the remaining bilinear terms in from the complementary slackness conditions \cref{eq:gen_comp_lower,eq:gen_comp_upper,eq:wind_comp_lower,eq:wind_comp_upper} using the standard Fortuny-Amat (``Big-M'') approach.

The resulting problem is the following MILP:
\allowdisplaybreaks
\begin{subequations}
\begin{align}
\min_{p_t^{\rm F}, p_t^{\rm w}, \lambda_t} \quad 
    & \sum_{t\in[T]}\Big(\sum_g c_g p_{g,t}^c - \lambda_t D_t + \sum_g \overline{\mu}_{g,t}^c \overline{p}_g \nonumber \hspace{-10cm}\\
    &\hphantom{\sum_{t\in[T]}\Big(} 
    + \sum_{i\in[I]}m_{i,t} K_t^{\rm var}\Big) + \sum_{d\in[D]}mp_{d} K^{\rm fix} \hspace{-10cm}\\
\text{s.t.} \quad 
    &\text{Wind farm constraints: \cref{eq:derated_capacity_bilevel,eq:maintenance_binary,eq:maintenance_requirement,eq:max_parallel_maintenance,eq:minimum_maintenance_time,eq:maintenance_start_indicator,eq:parallel_maintenance_indicator}} \hspace{-3cm}\\
    &\text{Dispatch stationarity: \cref{eq:kkt_bal_c,eq:kkt_bal_w}} \hspace{-3cm}\\
    &\text{Dispatch feasibility: \cref{eq:kkt_enerbal,eq:kkt_wind_primal_upper,eq:kkt_gen_primal_upper}} \hspace{-3cm}\\
    &\text{Dispatch variable domains: \cref{kkt:var_nonneg,kkt:var_free}} \hspace{-3cm}\\
    & \overline{\mu}_{g,t}^{\rm c} \leq F z_{g,t}^{\rm c,+}
    && \forall g,t, \\
    & p_t^{\rm F} A_t - p_t^{\rm w} \leq F(1-z_t^{\rm w,+})
    && \forall t, \\
    & p_{g}^{\rm c,max} - p_{g,t}^{\rm c} \leq F(1-z_{g,t}^{\rm c,+})
    && \forall g,t, \\
    & \underline{\mu}_{t}^{\rm w} \leq F z_t^{\rm w,-}
    && \forall t, \\
    & p_t^{\rm w} \leq F(1-z_t^{\rm w,-})
    && \forall t, \\
    & \underline{\mu}_{g,t}^{\rm c} \leq F z_{g,t}^{\rm c,-}
    && \forall g,t, \\
    & p_{g,t}^{\rm c} \leq F(1-z_{g,t}^{\rm c,-})
    && \forall g,t, \\
    & z_{g,t}^{\rm c,+},\ z_{g,t}^{\rm c,-},\ z_t^{\rm w,+},\ z_t^{\rm w,-} \in \{0,1\},
\end{align}%
\label{eq:bilevel_reformulated}%
\end{subequations}%
\allowdisplaybreaks[0]%
where we use $F$ as a large ``Big-M'' scalar instead of ``M'' to avoid confusion with our maintenance notation.

\section{Maintenance planning under uncertainty}
\label{sec:maint_planning_under_uncertainty}

We now consider uncertainty in the atmospheric and oceanic (``metocean'') conditions.
Specifically we consider that \textit{wind availability} and \textit{site access} is uncertain.
In contrast to traditional generator maintenance, offshore wind turbine maintenance has an additonal component of uncertainty because turbine access may be obstructed by unsafe metocean conditions, mainly wave height and wind speeds/gusts \cite{papadopoulos2024stochos}.
We model this uncertainty through a set of scenarios.

The set of wind availability scenario is $\{\bm{A}_{\omega}\}_{\omega=1}^{\Omega}$, where each scenario is the vector $\bm{A}_{\omega} = (A_{1,\omega},\dots,A_{t,\omega},\dots,A_{T,\omega})$.
Similarly, site access is the binomial variable $X_t\in\{0,1\}$ where, $X_t=1$ indicates that maintenance is possible at time~$t$.
Because the likelihood of the wind farm being accessible for maintenance is closely correlated with the atmospheric conditions that also determine wind power availability, we model site access as sets of scenarios conditional to each wind availability scenario $\omega$, i.e., $\{X_{t,\omega \nu}\}_{\nu=1}^{N_{\omega}},\ \forall \omega\in[\Omega]$ and $\bm{X}_{\omega \nu} = (X_{1,\omega \nu},\dots,X_{t,\omega \nu},\dots,X_{T,\omega \nu})$.
The joint scenario probability is $\pi_{\omega \nu} = \pi_{\omega}\pi_{\nu|\omega}$.

\subsection{Grid operator under uncertainty}
\label{ssec:grid_operator_under_uncertainty}

We assume that the grid operator does not make any look-ahead or stochastic decisions and computes the optimal cost-minimal dispatch for each scenario $\bm{A}_{\omega}$ once it is realized. 
In fact, any necessary look-ahead decisions, such as unit commitment status, can just be captured by the model as written. 
As a result, the economic dispatch problem remains the same as in \eqref{eq:ed_problem}, just with a change in notation such that all decision and dual variables receive an additional index $\omega$ to indicate their connection to a respective scenario. 
We write \cref{eq:ed_problem} as a function of scenario $\omega$ and the (derated) farm capacity as ${\rm ED}(\bm{A}_{\omega}, \bm{p}^{\rm F})$, where $\bm{p}^{\rm F} = (p_{1}^{\rm F},\dots,p_{t}^{\rm F},\dots,p_{T}^{\rm F})$.

\subsection{Strategic wind farm maintenance under uncertainty}
\label{ssec:strategic_osw_maintenance_under_uncertainty}

The wind operator only needs to decide the derated farm capacity $\bm{p}^{\rm F}$ (to report to the grid operator) and the maximum per-period number of parallel maintenance slots $mp_d$ (to procure equipment and crew) before the uncertainty is realized. 
The final actual maintenance schedule can be decided once uncertainty is resolved, i.e., actual wind availability and site access is known. 

The resulting strategic maintenance planning model under uncertainty is:
\allowdisplaybreaks
\begin{subequations}
\begin{align}
\max_{p_t^{\rm F}} \ 
    & \sum_{\omega\in[\Omega]}\!\!\pi_\omega\!\Bigg[\!\sum_{t\in[T]}\!\!\Big(p^{\rm w}_{t,\omega}\lambda_{t,\omega}\! - \!\!\!\sum_{\nu\in[N_\omega]} \pi_{\nu|\omega} \sum_{ i\in[I]}m_{i,t,\omega\nu} K_t^{\rm var}\big)\!\Big)\!\Bigg] \nonumber \hspace{-5cm}\\
    & -\sum_{d\in[D]}\!mp_{d} K^{\rm fix}  \\    
\text{s.t.} \ 
    & 0 \le p^{\rm F}_{t} \le \sum_{i\in[I]} \overline{p}_{i}(1-m_{i,t,\omega\nu}), && \forall i,t,\omega\nu \\
    & \sum_{t\in[T]}m_{i,t,\omega\nu} \ge M^{\rm req}_i &&\forall i, \omega\nu \\
    & ms_{i,t,\omega\nu} \ge m_{i,t,\omega\nu} - m_{i,t-1,\omega\nu} && \forall t, \omega\nu \\
    & \sum_{s=t-(M^{\rm int}_i-1)} ^t ms_{i,s,\omega\nu} \le m_{t,i,\omega\nu} && \forall t, \omega\nu \\
    & \sum_{i\in[I]}m_{i,t,\omega\nu} \le M^{\rm par}_t && \forall t,\omega\nu \\
    & mp_{d} \ge \sum_{i\in[I]} m_{i,t,\omega\nu} && \hspace{-2.5cm}\forall t \in [T]_d,\ \forall d\in[D],\ \forall \omega\nu \\
    & m_{i,t,\omega\nu} \in \{0,1\}, \ ms_{i,t,\omega\nu} \ge 0 && \forall i,t,\omega\nu \\
    & \lambda_{t,\omega}, p_{t,\omega}^w \in \argmin {\rm ED}(\bm{A}_{\omega}, \bm{p}^F) && \forall \omega \in [\Omega] \label{eq:stoch_inner}. 
\end{align}%
\label{eq:bilevel_uncertain}%
\end{subequations}%
\allowdisplaybreaks[0]%
For each $\omega$, the inner problem in \cref{eq:stoch_inner} allows the same reformulation as presented in Section~\ref{ssec:solution_approach}.
We highlight that the number of inner problems \eqref{eq:stoch_inner} scales only with $\Omega$ and is independent of $N_{\omega}$.

\section{Case Study}
\label{sec:case_study}

We use the derived models to study the role of strategic and grid-aware wind farm maintenance planning using the Grid Modernization Lab Consortium update of the IEEE Reliability Test System (RTS-GLMC) available at \cite{rts_glmc_git}.
Besides some hydro generation and solar PV assets, whose power injections we assume to be fixed by the given time-series data, the system hosts 73 conventional generators (nuclear, coal, gas and oil) and 3 large-scale wind farms. For this case study, we focus on the dataset's largest wind farm (``303\_WIND\_1''), which has an installed capacity of around 8\% of the system's peak load 16\% of the system's average load for the given one year data.
We acknowledge that this relative farm size is large. We study the impact of the relative farm size below but also remark that we do not model transmission and resulting congestion effects in this paper. 
In practice, such effects are likely to amplify the sensitivity of local prices to wind power availability, even for smaller relative wind farm sizes.

We first demonstrate some fundamental observations using the deterministic models presented in Section~\ref{sec:basic_model_formulation}.
We will use this model to study the impact of strategic maintenance planning on profits and electricity prices and analyze the impact of farm size and maintenance requirements.
We then focus on the uncertainty-aware models and study the impact of necessary pre-commitments and forecasting quality on wind farm profits and electricity prices.

The below numerical experiments have been implemented in Python and use Gurobi 13.0.1 with \textit{gurobipy} as solver and modeling language, respectively. 
All experiments were run on an Apple M5 CPU with 24GB memory. Our data and implementation is available open source at:
\vspace{-0.3em}
\begin{center}
\it
   \url{https://github.com/ropes-lab/osw-strategic-maintenance}.
\end{center}

\subsection{Illustrative deterministic case}
\label{ssec:illustrative_deterministic_case}

We first show an illustrative deterministic case of the effect of strategic maintenance planning. 
To this end, we model the studied wind farm to consist of 50 turbines, resulting in a per-turbine capacity of 15MW, which is a standard turbine size for modern offshore wind farms. 
We set $K^{\rm fix} = \$5000$ and $K^{\rm var} = \$500$ and enforce the following maintenance requirements: $M^{\rm req}_{i=1,...,10} = 20$, $M^{\rm req}_{i=11,...,20} = 10$, $M^{\rm req}_{i=21,...,30} = 5$. Further, we set $M_i^{\rm int} = 3,\ \forall i$ and $M_t^{\rm par} = 5,\ \forall t$.

\begin{figure}
    \centering
    \includegraphics[width=0.99\linewidth]{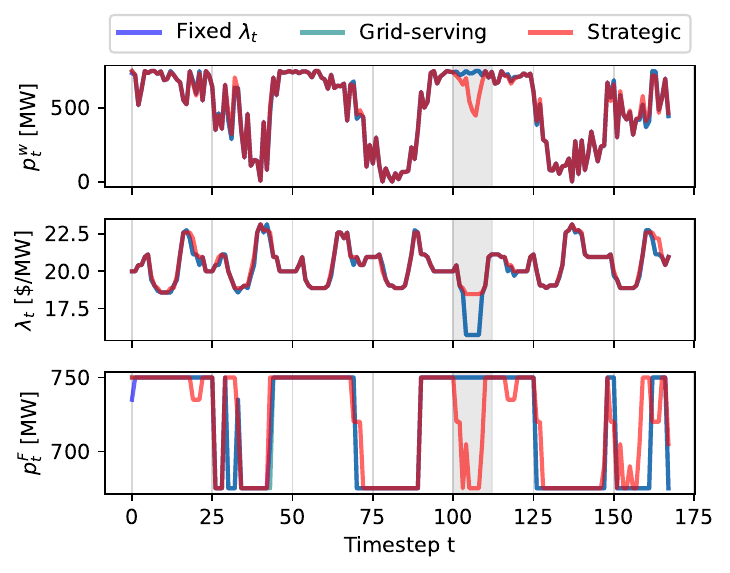}
    \caption{Injected wind power $p^{w}_t$, electricity price $\lambda_t$ and effective farm capacity $p^F_t$ over one week for maintenance planning with a fixed price $\lambda_t$ as per \eqref{eq:basic_maintenance_model}, grid-serving maintenance planning as per \cref{eq:grid_serving_maintenance}, and strategic maintenance planning as per \cref{eq:bilevel_reformulated}. 
    The shaded area highlights a period during which the strategic model deliberately reduces farm capacity despite high wind availability to avoid low prices and increase profits.}
    \label{fig:basic}
\end{figure}

We used this setup to solve the maintenance planning problem with fixed electricity prices as per \cref{eq:basic_maintenance_model}, a grid-serving maintenance planning model as per \cref{eq:grid_serving_maintenance}, and a strategic maintenance planning model as per \cref{eq:bilevel_reformulated}.
For for the basic approach, we first solved an economic dispatch problem as in \cref{eq:ed_problem} for one week at hourly resolution using the time series provided with the RTS-GLMC dataset starting on Jan. 1, 2020, and obtained prices $\lambda_t$ from this solution. 
We then used these prices to solve \cref{eq:basic_maintenance_model} and then re-ran the dispatch model with the reduced farm capacity to obtain the realized prices.
All maintenance planning models have been solved for the same time horizon and the results are shown in Fig.~\ref{fig:basic}.
The strategic model \cref{eq:bilevel_reformulated} solved to optimality (0.01\% gap) in 12 seconds.

We observe that the fixed-price and grid-serving approaches show very similar behavior leading to equally similar wind power injections, prices, and effective farm capacities $p^{\rm F}_t$. (Recall that a lower $p^{\rm F}_t$ indicates more maintenance intervals scheduled in parallel.)
In line with expected behavior from the opportunistic maintenance literature and validating our model, all three approaches aim to schedule maintenance during low-wind periods. For example, around hour 75. During these times, the opportunity cost of maintenance are much lower. 
However, at around hour 100, the fixed price and grid-serving models show a price drop caused by high wind availability during a lower load period. 
By scheduling some maintenance during this time and effectively removing some zero-cost wind production from the system, the strategic maintenance planning approach erases this price dip.
As a result, injected wind power is slightly reduced, but the remaining power sells at a higher price, which offsets the lost profits. 

\begin{figure}
    \centering
    \includegraphics[width=0.99\linewidth]{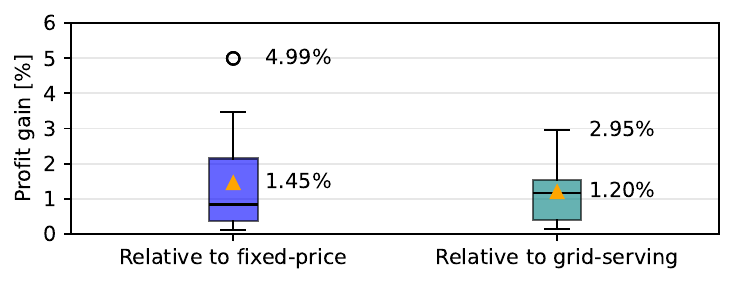}
    \caption{Profit gains per 10-day period relative to the fixed price and grid-serving maintenance planning approaches simulated over 30 periods with uniformly random starting times within a year of data. Box plots show the median as a line and mean as a triangle. The box covers results between the 25\% and 75\% percentile.}
    \label{fig:profits_det}
\end{figure}

To show the overall effect on profits we ran the same experiment for 30 ten-day periods picked randomly throughout the year of available data. This approach enables covering various seasonal effects while also giving us more variety from the available data. 
Fig.~\ref{fig:profits_det} shows the distribution of the profit gains of strategic maintenance planning relative to the fixed-price approach and grid-serving maintenance planning. 
The average profit increase is 1.45\% and 1.20\%, with maximum profit gains of up to 4.99\% and 2.95\%, respectively. 
Interestingly, the spread and average of the profit relative to the grid-serving approach are lower than compared to the basic, fixed-price approach. 
Our experiments showed that the grid-serving model sometimes deliberately schedules maintenance to avoid wind curtailment, thus avoiding some more extreme price drops and closing the gap to the strategic approach.

\subsection{Impact of farm and turbine size}

Next, still using the deterministic model, we analyze the impact of wind farm size and per-turbine power rating on the potential of the wind farm to gain additional profits from strategic maintenance planning. 
We set this analysis up as follows: 
We use the total capacity of the largest wind farm as given in the RTS-GLMC dataset and scale it by a value of 0.5 to 1.5 in 0.25 increments. 
We then compute the number of wind turbines by dividing the resulting scaled total farm capacity by a per-turbine rating of 12, 15, 18, or \unit[20]{MW} (rounding up). These ratings reflect current, planned, and prospective offshore wind turbine capacities \cite{musial2024scaling}.
To keep all cases comparable, we define maintenance requirements \textit{in percent} of the total farm capacity that match the illustrative case from Section~\ref{ssec:illustrative_deterministic_case}, i.e., the case with a farm scaler of 1.0 and a turbine size of \unit[15]{MW}. Hence, \unit[20]{\%} of the turbines require 20 hours, \unit[20]{\%} require 10 hours, and \unit[20]{\%} require 5 hours of maintenance.
The resulting relative farm sizes are \unit[4 to 12]{\%} of the system's peak load.
As before, we run each case for 30 random 10-day periods. 
We note that the numerical results are not identical for the \unit[15]{MW}-turbine at 1x scale-case relative to the results in Fig.~\ref{fig:profits_det} due to the resulting rounding in the turbine sizes and maintenance requirements.

\begin{figure}
    \centering
    \includegraphics[width=0.99\linewidth]{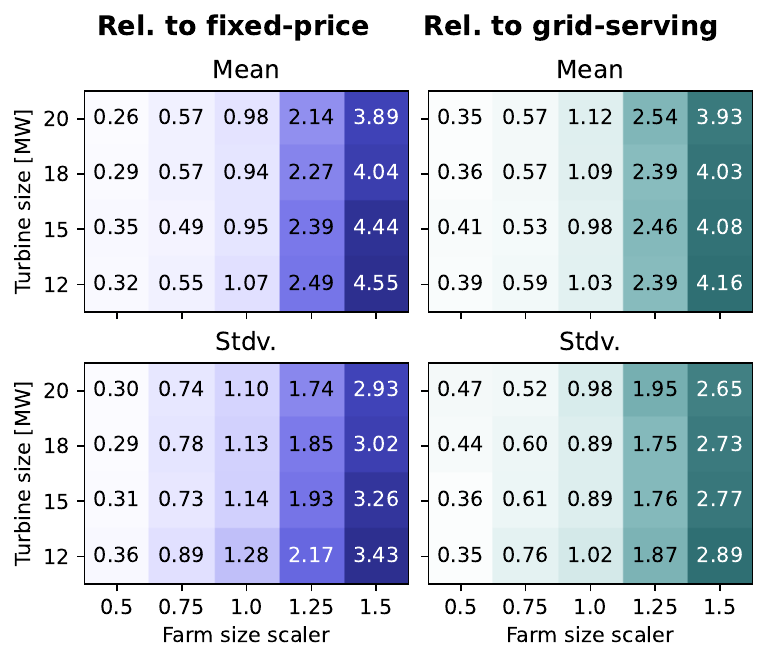}
    \caption{Means and standard deviations (stdv.) of relative profit gains for different farm sizes and per-turbine ratings. In each case, \unit[20]{\%} of the turbines require 20 hours, \unit[20]{\%} require 10 hours, and \unit[20]{\%} require 5 hours of maintenance.}
    \label{fig:profits_det_size_scale}
\end{figure}

Fig.~\ref{fig:profits_det_size_scale} shows the resulting profit gains relative to the fixed price maintenance scheduling and the grid-serving maintenance scheduling alongside their standard deviations.
We observe again that the overall profit gains relative to grid-serving are slightly lower relative to the fixed-price approach, because the grid-serving maintenance schedule avoids some very-low price periods. 
Overall, and as expected, the main driver of profit gains is the overall farm size because it increases the farm's impact on the system price. 
Interestingly, smaller turbine sizes correlate with higher profit gains which we explain with the fact that the wind farm operator has more flexibility to distribute required farm derates. 

\subsection{Maintenance planning under uncertainty}
\label{ssec:maintenance_planning_under_uncertainty}

We now analyze the impact of uncertainty on the strategic maintenance planning strategy and resulting profits using the model from Section~\ref{ssec:strategic_osw_maintenance_under_uncertainty}. 
To this end, we use a set of 10 wind availability scenarios $\bm{A}_{\omega}$ from \cite{papadopoulos2024stochos}, which reflect realistic, state-of-the-art forecasting techniques. 
For each scenario $\omega$ we estimate the access probability from the likelihood that wave height is higher than \unit[1.8]{m} \textit{or} that wind speed is higher than \unit[15]{m/s}.
Under these conditions offshore site access is generally considered unsafe, thus prohibiting maintenance \cite{papadopoulos2024stochos}. 
We use historical wind speed and wave-height data to estimate their relationship and the resulting likelihood that either the safety thresholds are exceeded using quadratic-kernel ridge regression.
We refer to the data and methods outlined in \cite{papadopoulos2024stochos} and its supplemental materials for details. For each wind scenario $\omega$, we sample $N_{\omega}=5$ sets of access scenarios.
The farm size and maintenance requirements correspond to the initial example from Section~\ref{ssec:illustrative_deterministic_case}.

\begin{figure}
    \centering
    \includegraphics[width=0.99\linewidth]{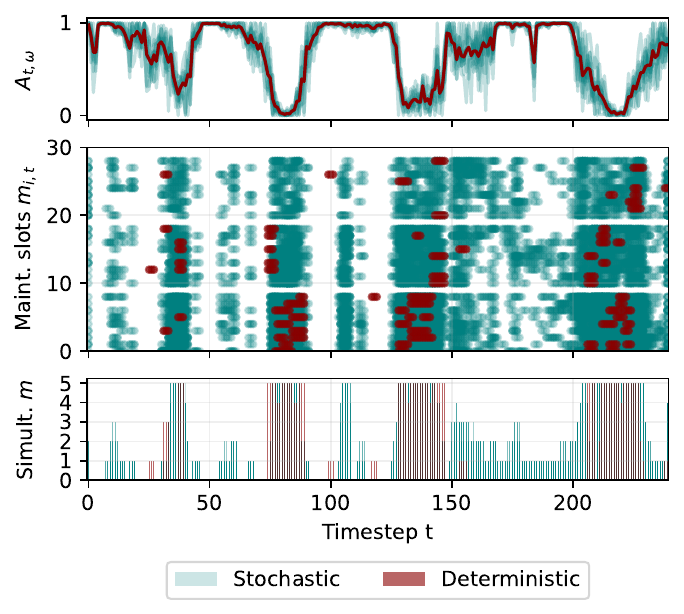}
    \caption{Comparison of strategic maintenance planning with uncertainty as per \eqref{eq:bilevel_uncertain} (in teal) and without uncertainty as per \eqref{eq:bilevel_reformulated} (in red). The top plot shows the wind availability scenarios. The middle plot shows the resulting maintenance slots overlaid for all scenarios. We observe them to be more spread out relative to the deterministic maintenance slots. Note that only turbine indices with non-zero maintenance requirements are shown.
    The bottom plot shows the total number of simultaneously scheduled maintenance slots.
    }
    \label{fig:stoch_access}
\end{figure}

With this setup, problem \eqref{eq:bilevel_uncertain} solved to a \unit[0.61]{\%} optimality gap in 23.2 minutes.
Fig.~\ref{fig:stoch_access} shows the wind scenarios and the resulting maintenance schedule of the stochastic planning problem in comparison to the deterministic result using the ensemble mean. 
The central observation in Fig.~\ref{fig:stoch_access} is that maintenance windows are now more spread out to account for the change in wind availability and resulting opportunity cost of maintenance.
On average over all scenarios, the model schedules an additonal \unit[3.46]{\%} of maintenance blocks to ensure that the maintenance targets are met.
As a result, profit decreases by \unit[1.76]{\%} relative to the perfect-knowledge deterministic baseline. 

\begin{figure}
    \centering
    \includegraphics[width=0.99\linewidth]{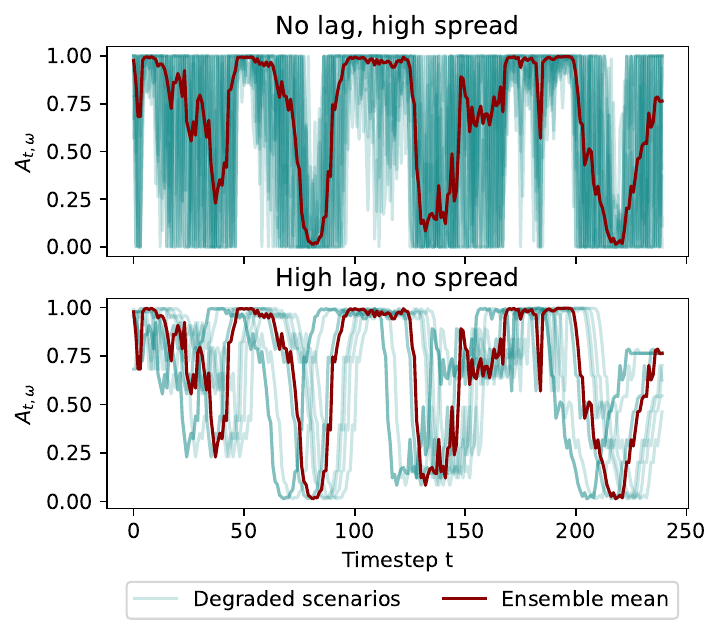}
    \caption{Illustration of forecast quality degradation by introducing spread (top) and lag (bottom). Spread increases the distance of each scenario from the ensemble mean, lag shifts each scenario relative to the ensemble mean.}
    \label{fig:forecast_degradation}
\end{figure}

Finally, we analyze the impact of forecast quality on the maintenance planning.
To this end, we artificially degrade the scenarios through two parameters: (i) The error level, which we call \textit{spread}, moves each scenario away from the ensemble forecast by a given factor. Formally, let $\bar{\bm{A}}$ be the forecast ensemble mean and define $\sigma$ to be the forecast spread factor. Then, the degraded forecast $\bm{A}_{\omega}'$ is:
\begin{equation*}
    \bm{A}_{\omega}' = \bar{\bm{A}} + \sigma(\bm{A}_{\omega} - \bar{\bm{A}}).
\end{equation*}
The final $\bm{A}_{\omega}'$ is clipped to the interval $[0,1]$ if needed. Note that the spread factor of the original forecasts is 1.
(ii) \textit{Lag} determines how scenarios are shifted against the ensemble mean, capturing the property of a forecast to be potentially very accurate in terms of the realized wind profile, but with a temporal shift. See also the discussion in \cite{ye2023airu}.
For each scenario we introduce lag by sampling a random integer number from $[-\ell,\ell]$, where $\ell\in\mathbb{Z}$ is the maximum number of timestep shifts. Realized, sampled lag can be positive of negative.
Fig.~\ref{fig:forecast_degradation} illustrates the effects of spread and lag on the scenarios.
The access uncertainty shares the degradation. 

Table~\ref{tab:forecast_quality_impact} shows the impact of forecast quality on the number of average additonal planned maintenance windows and changes in profit relative to the deterministic approach using the ensemble mean.
Across the cases shown in Table~\ref{tab:forecast_quality_impact}, model~\eqref{eq:bilevel_uncertain} solved in \unit[17.3]{minutes} on average to an average optimality gap of \unit[0.70]{\%}.

\begin{table}
\centering
\caption{Impact of forecast quality}
\label{tab:forecast_quality_impact}

\setlength{\tabcolsep}{3pt} 
\begin{subtable}[t]{0.45\linewidth}
    \centering
    \begin{tabular}{c c|c|c|c}
        & & \multicolumn{3}{c}{\textbf{Spread ($\sigma$)}} \\
        & & \textbf{0} & \textbf{1} & \textbf{10} \\
        \midrule
\multirow{3}{*}[-0.3em]{\rotatebox[origin=c]{90}{\textbf{Lag ($\ell$)}}}
        & \textbf{0} & 0.0 & 3.46 & 10.29 \\
        \cmidrule(l{0em}r{0.5em}){2-5}
        & \textbf{3} & 3.81 & 4.63 & 11.21 \\
        \cmidrule(l{0em}r{0.5em}){2-5}
        & \textbf{6} & 7.11  & 9.40 & 20.73
    \end{tabular}
    \vspace{0.5em}
    \caption{Additional maint. windows [\%]}
    \label{tab:forecast_quality_a}
\end{subtable}
\hfill
\begin{subtable}[t]{0.45\linewidth}
    \centering
    \begin{tabular}{c c|c|c|c}
        & & \multicolumn{3}{c}{\textbf{Spread ($\sigma$)}} \\
        & & \textbf{0} & \textbf{1} & \textbf{10} \\
        \midrule
\multirow{3}{*}[-0.3em]{\rotatebox[origin=c]{90}{\textbf{Lag ($\ell$)}}}
        & \textbf{0} & 0.0 & -1.76 & -11.11 \\
        \cmidrule(l{0em}r{0.5em}){2-5}
        & \textbf{3} & -2.06 & -0.85 & -10.43 \\
        \cmidrule(l{0em}r{0.5em}){2-5}
        & \textbf{6} & -2.56 & -1.02 & -10.75
    \end{tabular}
    \vspace{0.5em}
    \caption{Profit rel. to deterministic [\%]}
    \label{tab:forecast_quality_b}
\end{subtable}
\vspace{-2em}
\end{table}

Table~\ref{tab:forecast_quality_impact}(a) focuses on the additional number of maintenance windows that the wind farm operator schedules to ensure that maintenance targets can be met. 
We observe that at lower lags, the number of additonal maintenance windows is mainly driven by forecast uncertainty. 
At very high lags of up to six hours ($\ell=6$), the number of additionally scheduled maintenance windows more than doubles for each level of forecast uncertainty.
Interestingly, the impact on profit, see Table~\ref{tab:forecast_quality_impact}(b), is less monotonic as we observe that for $\sigma=0$ the profit loss is systematically higher across the studied lag cases than for $\sigma=1$. 
We attribute this to the fact that when forecast quality is only driven by lag, the model is less conservative in reducing farm capacity and ends up missing the relevant low-wind times, thus increasing its opportunity cost of maintenance. 
With slight forecast uncertainty, the resulting maintenance plan is more spread out, leading to better opportunities to schedule maintenance at low-opportunity cost windows. 
When forecast quality degrades further ($\sigma=10$), the resulting maintenance plan is very conservative, leading to strong increases in profit losses relative to the deterministic benchmark. 
Overall, we observe that profit losses are mainly driven by forecast uncertainty (i.e., spread), which we attribute to the fact that the system operator has full recourse for every scenario (see Section~\ref{ssec:grid_operator_under_uncertainty}).
Overall, we find that the economic impacts of the maintenance schedules are non-linear and non-monotonic in forecast quality, which opens opportunities for value-oriented forecasting approaches that jointly optimize prediction and decision-making.

\section{Discussion and Conclusion}

In this paper, we modeled and studied the maintenance planning problem for wind farms in a grid context using bilevel and stochastic programming approaches. 
To this end, we formulated a wind power maintenance planing problem alongside a market-clearing economic dispatch problem and formulated a bilevel program in which the wind farm operator makes a leading decision of derating its farm capacity to accommodate necessary turbine maintenance and the system operator then makes a following decision on how to price and dispatch generators to serve demand. 
We derived a deterministic and a stochastic version of this model, where the latter accounts for uncertainty in wind power availability and site access. 
We formulated a tractable solution approach for the models and used them to study the potential and mechanics of strategic, grid-aware maintenance planning. 

Our numerical experiments showed how wind farms can use low-price periods to reduce the opportunity cost of maintenance farm derates. At the same time our results show the potential of wind farms to act in a price-making capacity and avoid very-low price periods by scheduling maintenance of some turbines strategically during high-wind, low-demand periods. 
Interestingly, this objective is partially related to a fully grid-serving maintenance plan selected by the system operator. 
As expected, our results showed that uncertainty in wind power availability and access to the site to successfully perform maintenance degrades the profit margin and requires scheduling additonal maintenance intervals to ensure that all maintenance requirements can be fulfilled. 

Some details of our modeling choice are worth discussing here. 
First, we assumed that the leader decision (wind farm operator) has perfect information on the follower's behavior (the system operator). 
Hence, our results are an upper bound on the additonal profits that strategic and grid-aware maintenance planning can achieve within the context of our experiments. 
In practice, the wind farm operator has to perform additonal prediction tasks for grid signals and potentially robustify their maintenance planning schedule against uncertainty with respect to these signals. 
Forecasting such grid signals and making maintenance decisions based on them is an emerging research direction that we also explore in ongoing work~\cite{wang2026gridinformed,galsim2026curtailment}. We also  highlight the opportunity for task- or value-oriented forecasting (see, e.g., \cite{ghazanfariharandi2025value}) as discussed above.

Second, we simplify practical market clearing and assumed no grid congestion. 
Incorporating a detailed market clearing with power flow constraints into the follower decision will complicate the model and will likely require more elaborate solution techniques, e.g., as in \cite{mieth2024prescribed,ghazanfariharandi2026uncertainty}.
Regardless, and as mentioned in Section~\ref{eq:grid_serving_maintenance} above, the presence of grid congestion is expected to amplify the ability of the wind farm to influence prices, which lead us to conclude that this assumption is not critical for the results presented in this paper. 

Finally, we focused on a single wind farm. In practice, multiple offshore wind farms might be connected to grid connection points that are electrically close in the network. 
This motivates further research on the impact of multiple strategic wind farm operators by extending our modeling approach to a multi-leader, single-follower game. 
In this context, it would also be worthwhile to study the effects of other shared markets, e.g., for maintenance vessels and parts, that these wind farm operators act in. 
We consider this line of research to be beyond the scope of this paper and will consider it in future research.

\section*{AI Disclosure}

For the preparation of this work, the authors used generative AI tools to support some coding implementation grammar and spelling.
After any AI use, the authors reviewed and edited the content and take full responsibility for the the article. No reasoning or results analysis was done by AI and no AI has been used on the references.

\balance
\bibliographystyle{IEEEtran}
\bibliography{literature}

\end{document}